\documentclass[epj,final]{svjour}

\usepackage{epsfig}
\usepackage{graphicx}
\usepackage{amsmath,amsfonts}

\newcommand{\bs}{\boldsymbol}

\newcommand{\ml}{\mathcal}

\newcommand{\ket}{\rangle}
\newcommand{\bra}{\langle}

\begin{document}
\title{Partonic transverse momenta in non relativistic hyper-central quark potential models}
\author{F.K. Diakonos, N.K. Kaplis, X.N. Maintas}
\institute{University of Athens, Physics Department, Section of Nuclear and Particle Physics, GR-15771, Athens, Greece\\ \email{fdiakono@phys.uoa.gr}}
\authorrunning{F.Diakonos,N.Kaplis,X.Maintas}
\titlerunning{Partonic transverse momenta in hyper-central quark potential models}

\date{Received: 15-01-2009 / Revised version: 24-03-2009}

\abstract{
We investigate the impact of three-body forces on the transverse momentum distribution of partons inside the proton. This is achieved by considering the three body problem in a class of hyper-central quark potential models. Solving the corresponding Schr\"odinger equation we determine the quark wave function in the proton and with appropriate transformations and projections we find the transverse momentum distribution of a single quark. In each case the parameters of the quark potentials are adjusted in order to sufficiently describe observable properties of the proton. Using a factorization ansatz we incorporate the obtained transverse momentum distribution in a perturbative QCD scheme for the calculation of the cross section for prompt photon production in $pp$ collisions. A large set of experimental data is fitted using as a single free parameter the mean partonic transverse momentum. The dependence of $\langle k_T \rangle$ on the collision characteristics (initial energy and transverse momentum of the final photon) is much smoother when compared with similar results found in the literature using a Gaussian distribution for the partonic transverse momenta. Within the considered class of hyper-central quark potentials the one with the weaker dependence on the hyper-radius is preferred for the description of the data since it leads to the smoothest mean partonic transverse momentum profile. We have repeated all the calculations using a two-body potential of the same form as the optimal (within the considered class) hyper-central potential in order to check if the presence of three body forces is supported by the experimental data. Our analysis indicates that three-body forces influence significantly the form of the parton transverse momentum distribution and consequently lead to an improved description of the considered data.
}
\maketitle

\section{Introduction}
The description of the hadron spectra is still an open question in theoretical physics. To date, the most important progress in this direction is based on lattice QCD, QCD sum rules and potential models. Despite being more fundamental, lattice QCD and QCD sum rules can only lead to rough description of hadronic spectra. Potential models on the other hand, albeit not fundamental, have been proved to be very successful even in the non-relativistic approximation.

Since the late seventies several attempts have been made in this field using different forms for the quark-quark interacting potential, leading in many cases to a very good description of the hadronic states. The main ingredient in all these models is the presence of a confining part in the inter-quark potential, which is independent of flavor and spin.

Although in light baryons, as the proton, relativistic effects are expected to play a role, there are several non-relativistic treatments leading to satisfactory description of the proton properties \cite{Martin81,Richard81,QuarkModels,Reyes03}. In particular the non-relativistic inter-quark potential:
\begin{equation}
V(r) = A + B r^{0.1}
\label{eq:eq1}
\end{equation}
with $A$, $B$ appropriate constants, has been successfully used in the literature for the  description of the heavy quark meson wave function as well as the clearly relativistic $s\bar{s}$ states \cite{Martin81}. The same model was used later in \cite{Richard81} in order to obtain baryonic spectra with very good results. With the progress of lattice QCD it became possible to determine effective inter-quark potentials based on first principles \cite{Hasenfratz80,Richard92}. During the late eighties it was realized that genuine many body interactions could also play an important role in the determination of baryon properties \cite{Capstick86}. To this end it has been proved to be very efficient to express the interaction between the quarks in terms of hyper-radial potentials \cite{Ferraris95,Santopinto98,Giannini01,Giannini02}. In this treatment $V(\xi)$ is in general a three body potential since the hyper-radius $\xi$ depends on the coordinates of all three particles. Such potentials have been extensively used for a consistent description of a large set of hadronic observables which besides their spectra include the photo-couplings \cite{Aiello96}, the electromagnetic form factors and the strong decay amplitudes \cite{Strobel96,Chen07}. 
Recently it has been proposed that within the potential model approach one could also obtain an estimation of the transverse momentum distribution of partons inside the hadrons $g(k_T)$ \cite{PRD1}. The resulting probability density, characterized by a non-Gaussian shape, was then used within the framework of perturbative QCD for a phenomenological description of the cross section for $\pi^0$ production in $pp$ collisions. Interestingly enough, this treatment turned out to be very efficient in the description of the experimental data resolving several unsatisfactory issues present in the usual approach involving a Gaussian form for $g(k_T)$. In \cite{PRD1} the treatment was based on a two-body potential of the form (\ref{eq:eq1}) while in a later work the MIT bag model \cite{MITbag} was used in a similar manner to obtain $g(k_T)$ and subsequently to describe successfully the prompt photon production in $pp$ collisions \cite{PRD2}. The results of these two works indicate that confinement, asymptotic freedom and/or relativistic description make an imprint on the intrinsic transverse momentum distribution of the constituent quarks, detectable in the cross section of $pp$ collisions. As mentioned in \cite{PRD2} this fact could give an explanation for the systematic discrepancy between theoretical next-to-leading order (NLO) calculations \cite{Florian05} and experimental data \cite{Apanasevich98,Ballocchi98} of inclusive single photon production ($pp \to \gamma X$). The observed gap for this process is particularly significant in fixed target experiments and cannot be filled even after taking into consideration certain large contributions to the partonic hard scattering cross section to all orders in perturbation theory, using the threshold resummation technique \cite{Sterman87,Laenen98,Catani99}. 

In the present work our main interest is to explore if genuine three-body effects may also influence the transverse momentum distribution of the quarks inside the proton in a way that it is detectable in cross section data. Therefore, we initially attempt a consistent description of the ground state wave function of the proton as a three quark bound state within a class of hyper-central quark potential models of the form:
\begin{equation}
V(\xi) = A_k + B_k  \xi ^k
\label{eq:eq2}
\end{equation}
where $A_k$, $B_k$ are constants, $k \in R^+$ and $\xi$ is the hyper-radius. Due to the many body nature of the problem in the general case the wave functions can only be obtained numerically. One interesting exception is when the quark-quark confining potential is harmonic, allowing for analytical solutions. In addition, the harmonic model supplies a convenient classification scheme of the baryon resonances in terms of shells 
\cite{Reyes03}. The parameters $A_k$ and $B_k$ are chosen in order to fit the proton's ground and first excited state energy. To ensure consistency we also estimate the proton's charge radius. Having fixed $A_k$, $B_k$ we determine $g(k_T)$ for several values of $k$ ($k=0.1,0.5,1,2$) and then we use the standard treatment within perturbative QCD for the calculation of the cross section for prompt photon production in a $pp$ collision experiment. We compare our results with those of \cite{PRD1,PRD2} in order to extract information concerning the presence or not of traces of three-body effects in $g(k_T)$ traceable through the considered experimental data.

Our work is organized as follows. In section $2$ we introduce the hyper-central description of the three body problem in the non-relativistic case using the potential $V(\xi) = A_k + B_k \xi^k$. We solve numerically the hyper-radial Schr\"odinger equation for the 4 different values of 
$k$ mentioned above. In each case we determine the parameters $A_k$ and $B_k$ having as criterion the exact description of the ground and first excited state energy of the proton. In section $3$ we calculate the intrinsic transverse momentum distribution $g(k_T,\bra k_T \ket)$ of partons inside the proton for different $k$. In section $4$ we present the numerical results from the best fit of the $pp \to \gamma X$ cross section data at various energies and transverse momenta of the produced photon, using the single parameter distribution $g(k_T , \bra k_T \ket)$ within the framework of perturbative QCD. Finally, section $5$ contains our concluding remarks.

\section{Hyper-central potential for the proton}

We start our study by considering the proton as a bound state of three constituent quarks. After fixing the center of mass, the three particle configuration is described by the Jacobi coordinates:
\begin{eqnarray}
 \bs{\xi}_1 = \bs{x}_2 - \bs{x}_1 \nonumber\\
 \bs{\xi}_2 = \frac{2\bs{x}_3 - \bs{x}_1 - \bs{x}_2}{\sqrt{3}}
 \label{eq:eq3}
\end{eqnarray}
Instead of $\bs{x}_i$ ($i=1,2,3$) one can introduce the hyper-spherical coordinates, which are given by the angles $\Omega_1 = (\theta_1, \phi_1)$ and $\Omega_2 = (\theta_2 , \phi_2)$ ($\theta_i$, $\phi_i$ are polar and azimuthal angles of vector $\bs{\xi}_i$, $i=1,2$) along with the hyper-radius $\xi$ and the hyper-angle $\chi$, defined by the relations:
\begin{eqnarray}
\xi_1 = \xi \cos\chi\nonumber\\
\xi_2 = \xi \sin\chi
\label{eq:eq4}
\end{eqnarray}
In this model we consider three identical quarks of mass $m$. Then the Hamiltonian can be written as:
\begin{equation}
H = -\frac{\hbar^2}{2\mu}\left( \nabla_{\xi_1}^2 + \nabla_{\xi_2}^2 \right) + V(\xi)
\label{eq:eq5}
\end{equation}
where $\mu = \frac{m}{2}$ and the potential depends only on $\xi$ (hyper-central). Since $\xi^2 = \xi_1^2 + \xi_2 ^2$ the interaction in Eq. \ref{eq:eq5} is not a purely two body interaction but contains three body contributions. The presence of three quark forces could be suggested by the existence of a direct three gluon interaction which is one of the fundamental features of the non-abelian nature of QCD. In fact all these many body terms can be included, effectively, in an appropriate hyper-central potential \cite{Giannini02}.

\begin{figure}
\includegraphics[width=9. cm]{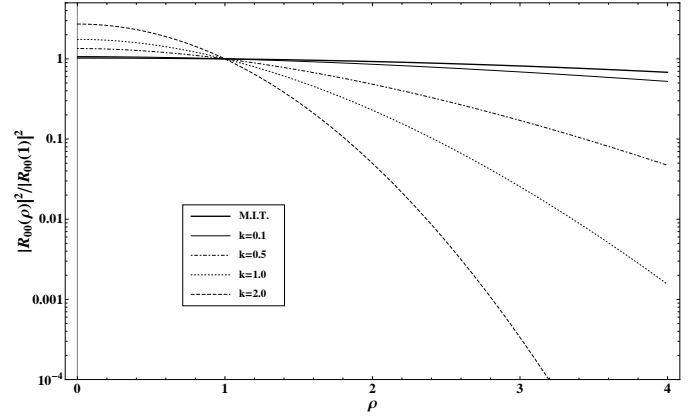}
\caption{The ground state wave function of the proton using the hyper-radial potential (\ref{eq:eq2}) for four different choices of $k$:
0.1 (solid line), 0.5 (slashed-dotted line), 1.0 (dotted line), 2.0 (slashed line) for the potential $V(\rho)=\alpha_k + \beta_k \rho^k$. The thick solid line correspond to the MIT bag model. The inset displays the ground state wave function for $k=0.1$ in an appropriate scale.}
\label{fig:fig1}
\end{figure}

\begin{figure}
\includegraphics[width=9. cm]{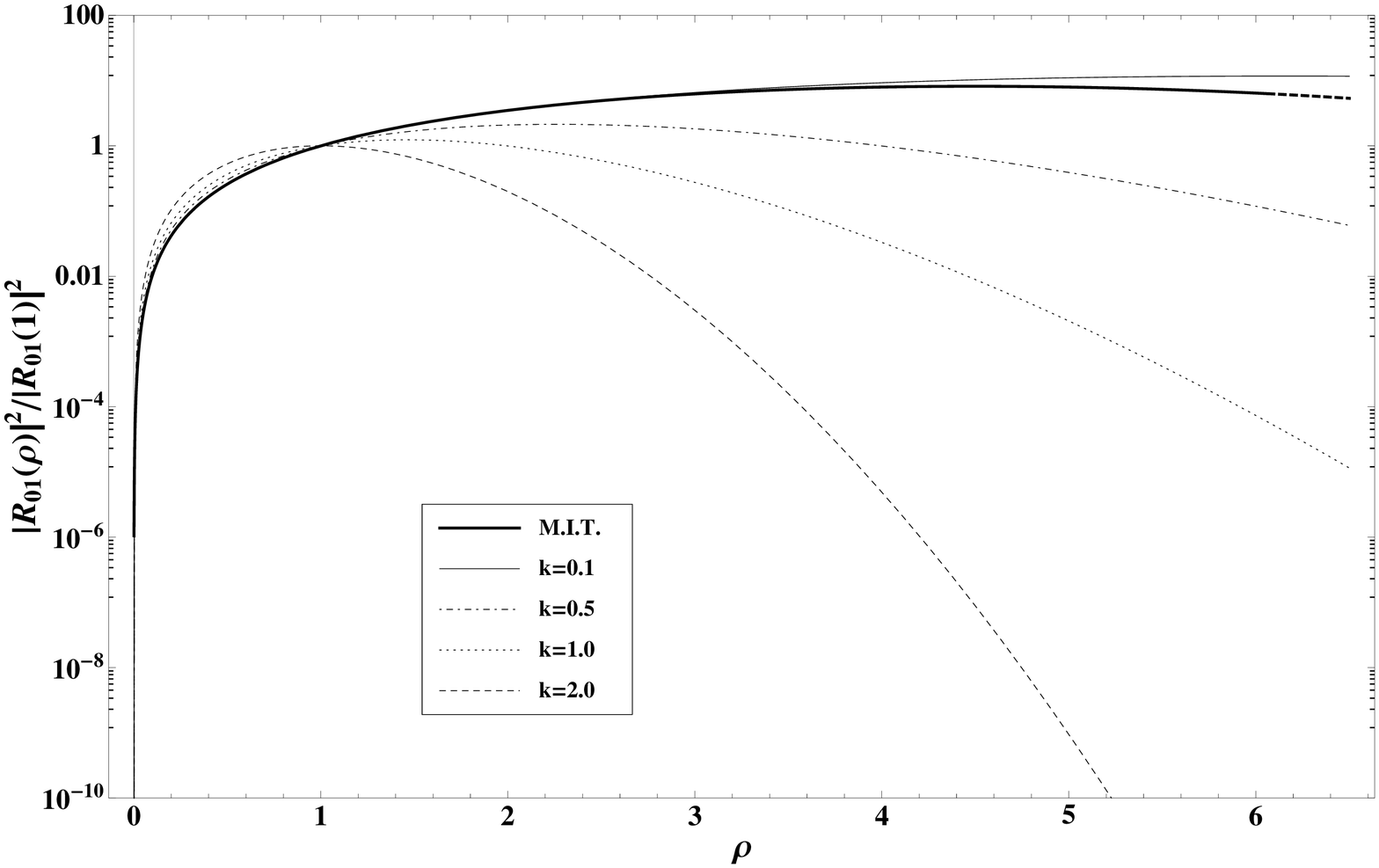}
\caption{The first excited state wave function of the proton using the hyper-radial potential (\ref{eq:eq2}) for four different choices of $k$:
0.1 (solid line), 0.5 (slashed-dotted line), 1.0 (dotted line), 2.0 (slashed line) for the potential $V(\rho)=\alpha_k + \beta_k \rho^k$. The thick solid line correspond to the MIT bag model. The inset displays the ground state wave function for $k=0.1$ in an appropriate scale.}
\label{fig:fig2}
\end{figure}

Using hyper-spherical coordinates the Hamiltonian of the three body problem can be written as:
\begin{equation}
H = - \frac{1}{2\mu}\left[ \frac{1}{\xi^2}\frac{\partial}{\partial \xi} \left( \xi^2 \frac{\partial}{\partial \xi} \right) + \frac{1}{\xi^2} 
\ml{L}^2 (\Omega_1,\Omega_2,\chi)\right] + V(\xi)
\label{eq:eq6}
\end{equation}
where:
\begin{equation}
\ml{L}^2(\Omega_1,\Omega_2,\chi) = \frac{1}{\sin^22\chi}\frac{\partial}{\partial \chi}\left( \sin^2 2\chi \frac{\partial}{\partial \chi} \right) + \frac{J^2}{\cos^2\chi} + \frac{L^2}{\sin^2 \chi}
\label{eq:eq7}
\end{equation}
in which $\hat{J}$ is the angular momentum of the subsystem of particles $1,2$ and $\hat{L}$ is the angular momentum of particle $3$ with respect to the center of mass of the two body subsystem ($1,2$). $\ml{L}^2(\Omega_1,\Omega_2,\chi)$ is the Casimir operator of the six dimensional rotation group $O(6)$ and its eigenfunctions are the hyper-spherical harmonics $Y_{\lambda j m_j l m_l}(\Omega_1,\Omega_2,\chi)$. That is:
\begin{eqnarray}
\ml{L}^2(\Omega_1,\Omega_2,\chi) Y_{\lambda j m_j l m_l}(\Omega_1,\Omega_2,\chi)= \nonumber \\
= -\lambda(\lambda +4)Y_{\lambda j m_j l m_l}(\Omega_1,\Omega_2,\chi)
\label{eq:eq8}
\end{eqnarray}
where the grand-angular quantum number $\lambda$ is given by $\lambda = 2n + j + l$, $n=0,1,2,\ldots$ and $j$, $l$ are the angular momenta associated with the $\hat{J}$ and $\hat{L}$ operators respectively. The solution of the Schr\"odinger equation in this case has the following form:
\begin{eqnarray}
&\Psi&_{N\lambda j m_j l m_l}(\xi , \chi, \theta_1, \phi_1, \theta_2,\phi_2)=\nonumber \\
 &=& N_{N\lambda j m_j l m_l} R_{N\lambda}(\xi) Y_{\lambda j m_j l m_l}(\theta_1,\phi_1,\theta_2,\phi_2,\chi) \phantom{aaa}
\label{eq:eq9}
\end{eqnarray}
where
\begin{eqnarray}
Y_{\lambda j m_j l m_l}(\theta_1&,&\phi_1,\theta_2,\phi_2,\chi)= \cos^{j + \frac{1}{2}}\chi \sin^{l + \frac{1}{2}}\chi \ml{P}_{n}^{l+\frac{1}{2}, j + \frac{1}{2}}\cdot \nonumber \\
&\cdot& (\cos2\chi) Y_{j}^{m_j}(\theta_1,\phi_1)Y_{l}^{m_l}(\theta_2,\phi_2)\phantom{a}
\label{eq:eq10}
\end{eqnarray}
and $\ml{P}_{n}^{l + \frac{1}{2}, l + \frac{1}{2}}(\cos2\chi)$ are the Jacobi polynomials.
For the hyper-radial many-body interaction between equal mass particles we use the general form (\ref{eq:eq2}) leading to the radial equation:
\begin{eqnarray}
\left[ - \frac{1}{\mu}\frac{1}{\xi^5} \frac{\partial}{\partial \xi} \left( \xi^5 \frac{\partial}{\partial \xi} \right) \right. &+& \left.  \frac{\lambda(\lambda + 4)}{2\mu \xi^2} + B_k \xi ^k - \left(  E_{N\lambda} - A_k \right)\right] \cdot \nonumber \\
&\cdot& R_{N\lambda}(\xi)=0 
\label{eq:eq11}
\end{eqnarray}
Introducing new variables
\begin{eqnarray}
&&\rho = \sqrt{\beta} \frac{\xi}{\xi_0} \ , \ \xi_0= \left( \frac{1}{2\mu \ml{\epsilon}} \right)^{\frac{2}{k+2}} \ , \ \ml{\epsilon} = E_{N\lambda} - A_k \nonumber \\
&&\beta^{\frac{k + 2}{2}}= \frac{B_k}{2\mu \ml{\epsilon}^2} \ , \ W=2\mu \xi_0 ^2 \beta^{-1}
\label{eq:eq12}
\end{eqnarray}
and the new function:
\begin{equation}
u_{N\lambda}(\rho) = \rho^{5/2} R_{N\lambda}(\rho)
\label{eq:eq13}
\end{equation}
As a result Eq.~(\ref{eq:eq11}) becomes:
\begin{equation}
\left[ \frac{d^2}{d\rho^2} - \frac{(\lambda + \frac{3}{2})(\lambda + \frac{5}{2})}{\rho^2} - \rho^k + W\ml{\epsilon} \right]u_{N\lambda}(\rho) = 0
\label{eq:eq14}
\end{equation}
determining the energy eigenvalue problem to be solved. As discussed previously the non trivial part of the calculation is the solution of the radial equation (\ref{eq:eq14}), which in the general case (arbitrary $k$) can be obtained only numerically using the Numerov algorithm. Actually, solving (14) we determine simultaneously the eigenvalues $\sigma_0= W \epsilon_0$ and $\sigma_1 = W \epsilon_1$ and using the equations:
\begin{eqnarray}
\tau (\mu , B_k) A_k = \tau (\mu , B_k) E_i - \frac{2 \sigma_i}{E_i } \nonumber \\
i=0,~1 \ ; \ \tau (\mu , B_k) = 2 \mu \frac{\xi_0^4}{\beta^2}
\label{eq:eq15}
\end{eqnarray}
where $E_0=0.938~GeV$ and $E_1=1.440~GeV$ are the proton's ground and first excited state energies respectively, we find the constants $A_k$ and $B_k$ of the hyper-radial potential in (\ref{eq:eq2}). These quantities are necessary in order to get estimations of dimensionfull observables from our calculations. In Figs.~\ref{fig:fig1},\ref{fig:fig2} we present the wave functions of the ground state and the first excited state respectively (properly normalized for dimensional reasons) for the four different values of $k$ mentioned above. The insets display in more detail, using a suitable scale, the form of the ground state (Fig.~\ref{fig:fig1}) and the first excited state (Fig.~\ref{fig:fig2}) for $k=0.1$. Using the ground state wave functions shown in Fig.~\ref{fig:fig1} we can calculate the hyper-radius $\langle \xi^2 \rangle^{\frac{1}{2}} = \frac{\xi_0^2}{\beta} \langle \rho^2 \rangle^{\frac{1}{2}}$ of the proton ($\langle \rho^2 \rangle^{\frac{1}{2}}$ being the corresponding dimensionless quantity).

The hyper-radius can be related with the experimentally accessible charge radius $r_{ch}=\sqrt{\langle \bs{x}_i^2 \rangle}~,~~~i=1,2,3$ (assuming $\langle \bs{x}_1^2 \rangle = \langle \bs{x}_2^2 \rangle =\langle \bs{x}_3^2 \rangle$) through: $\sqrt{\langle \bs{x}_i^2 \rangle}_{ch}=\frac{1}{2} \sqrt{\langle \xi^2 \rangle}$. In Table I we summarize the results for the proton charge radius using the four different values of $k$ previously mentioned.\\

\begin{table}[h]
\caption{\label{tab:table1} The proton charge radius for different choices of the exponent $k$ in the hyper-radial potential.}
\begin{center}
\begin{tabular}{c|c}
k    & $r_{ch}$ (fm) \\ \hline
0.1  & 0.60343  \\ 
0.5  & 0.60695  \\
1    & 0.60901  \\
2    & 0.61002  \\
\end{tabular}
\end{center}
\end{table}

According to Table I, based on the value of the proton's charge radius no distinction between the four considered cases of $k$ is possible.

\section{Intrinsic Transverse Momentum Distribution $g(k_T, \bra k_T \ket)$}

As a next step we determine, for each choice of $k$, the single particle transverse momentum distribution in the ground state. In order to proceed we first have to calculate the ground state wavefunction in the momentum space (conjugate to the space $\bs{\xi}_i \ , \ i=1,2,3$) as:
\begin{eqnarray}
\tilde{\phi}(k_\xi) = N \int d\xi d\chi d\Omega_1 d\Omega_2 \xi^5 \cos^2 \chi \sin^2 \chi \cdot \nonumber \\
\cdot \exp[-\imath k_\xi \xi \cos \chi \cos\theta_1] \Psi_{N\lambda j m_j l m_l}(\bs{\xi})\phantom{aaa}
\label{eq:eq16}
\end{eqnarray}
where $\Psi_{N\lambda j m_j l m_l}$ is the full eigenfunction. A convenient representation of the integrals in eq.~(\ref{eq:eq16}) is achieved in the reference frame where $\bs{k}_\xi = (0,0,k_\xi,0,0,0)$.

\begin{figure}
\includegraphics[width=9. cm]{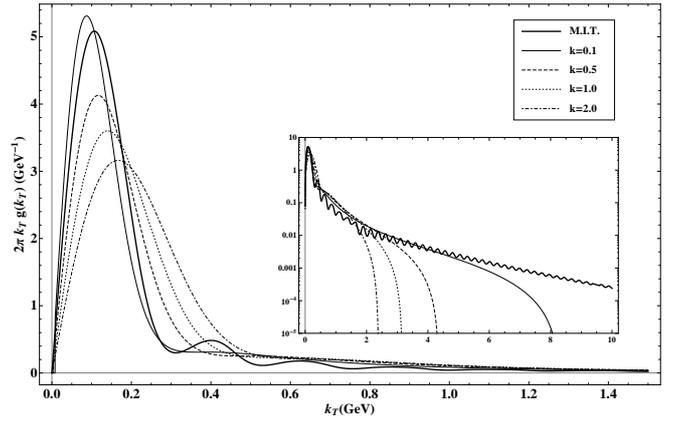}
\caption{Transverse momentum distribution of a parton inside the proton corresponding to the different values of $k$ in the hyper-radial  potential model (\ref{eq:eq2}). The thick solid line corresponds to the MIT bag model.}
\label{fig:fig3}
\end{figure}

It is useful to determine the transformation of the momenta $\bs{k}_{\xi_i}$ to the Cartesian momenta $\bs{k}_i$:
\begin{eqnarray}
\bs{k}_{\xi_1} = -\frac{1}{2}(\bs{k}_1 - \bs{k}_2) \nonumber \\
\bs{k}_{\xi_2} = -\frac{1}{2\sqrt{3}} (\bs{k}_1 + \bs{k}_2 - 2\bs{k}_3) \nonumber \\
\bs{k}_{\xi_3} = \frac{1}{\sqrt{3}}(\bs{k}_1 + \bs{k}_2 + \bs{k}_3)
\label{eq:eq17}
\end{eqnarray}
The above expressions is simplified, in the center of mass frame where $\bs{k}_{\xi_3}=0$ and:
\begin{equation}
k_\xi ^2 = k_1^2 + k_2 ^2 + \bs{k}_1 \cdot \bs{k}_2
\label{eq:eq18}
\end{equation}
The two particle density $\rho(\bs{k}_1,\bs{k}_2)$ is then given by:
\begin{equation}
\rho(\bs{k}_1,\bs{k}_2) = \left| \tilde{\phi}(k_\xi) \right|^2 = \left| \tilde{\phi}(\sqrt{k_1^2 + k_2 ^2 + \bs{k}_1 \cdot \bs{k}_2}) \right|^2
\label{eq:eq19}
\end{equation}
Finally from Eq. \ref{eq:eq19} we obtain the one particle transverse momentum density $g(k_T)$ as:
\begin{eqnarray}
g(k_T) = 4\pi \int_{-\infty}^{+\infty} dk_z \int _{-1}^{+1} dz \int_{0}^{+\infty} dk_2 k_2 ^2 \cdot \nonumber \\
\cdot \left| \tilde{\phi}
(\sqrt{k_T^2 + k_z^2 + k_2 ^2 + z k_2 \sqrt{k_T ^2 + k_z^2}}) \right|^2 \phantom{aa}
\label{eq:eq20}
\end{eqnarray}
with $z=\cos \omega$ where $\omega$ is the angle between $\bs{k}_1$ and $\bs{k}_2$.
The integration in equations \ref{eq:eq16} and \ref{eq:eq20} can be performed to a great accuracy using a mixture of Gauss-Kronrod quadrature and VEGAS Monte-Carlo integration algorithm. In Fig.~\ref{fig:fig3} we present the intrinsic transverse momentum distribution $2\pi k_T g(k_T)$ of a parton inside the proton obtained from the ground state wave function corresponding to each of the four different choices of the exponent $k$ in (\ref{eq:eq2}).
As expected for increasing $k$ the maximum of the distribution becomes broader while the tail tends to be more abrupt.

One possibility for testing the phenomenological relevance of three-body forces is to consider their influence in the description of physical processes, like prompt photon production in $pp$-collisions, within the framework of perturbative QCD. Our strategy is to find the value of $k$ in (\ref{eq:eq2}) for which we achieve the best description of experimental data and then to compare our results with those obtained using $g(k_T)$
determined through the MIT bag model \cite{PRD2} or the corresponding two-body potential \cite{PRD1}.

\section{Numerical Results}

We consider prompt photon production in $pp$-collisions as the appropriate process for checking the influence of three-body partonic interactions in the phenomenology of proton collisions. In fact this process is optimal for this purpose since it is not affected by experimental ambiguities caused by final state hadronic interactions. In order to proceed we use the phenomenological scheme proposed in \cite{PRD1,PRD2} for the calculation of the differential cross-section for inclusive $\gamma$-production. This scheme incorporates partonic subprocesses according to perturbative QCD, partonic effects in the proton described through the longitudinal parton distribution functions (PDF) and effects due to the intrinsic transverse momenta of the partons described through $g(k_T)$. A simplified phenomenological approach is adopted, in which it is assumed a factorization between longitudinal and transverse momentum parton distributions \cite{Feynman77,Owens87}. Although such an assumption seems reasonable from a statistical point of view, since the longitudinal momenta of the partons may differ by orders of magnitudes from the corresponding transverse ones, its validity based on first principles remains under question \cite{Bomhof:2007xt,Collins:2007nk}. Despite this fact this factorization asatz turned out to work sufficiently well in the case of cross section calculations for prompt photon production using the MIT bag model \cite{PRD2}. Here, as we are interested in comparing results obtained using different inter-partonic interactions for the description of the proton wave function, it is necessary to use exactly the same treatment as that introduced in \cite{PRD1,PRD2}. The calculations are performed in next-to-leading order (NLO) of perturbative QCD and the cross section for single photon production is given by: 
\begin{eqnarray}
E_{\gamma} \frac{d^3 \sigma}{d^3 p}(pp \rightarrow \gamma + X)=
K(p_T,\sqrt{s}) \sum_{abc} \int dx_a dx_b \cdot\nonumber \\
\cdot f_{a/p}(x_a,Q^2) f_{b/p}(x_b,Q^2)\cdot \frac{\hat{s}}{\pi}\frac{d \sigma}{d
\hat{t}}(ab \rightarrow c \gamma) \delta(\hat{s}+\hat{t}+\hat{u})\ \ \ \  
\label{eq:eq21}
\end{eqnarray}
where $f_{i/p}$ ($i=a,b$) are the MRST2006 NNLO longitudinal parton distribution functions (PDF) for the colliding partons $a$ and $b$ as a function of longitudinal momentum fraction $x_i$ and factorization scale $Q$ \cite{MRST2006}. $\frac{d \sigma}{d \hat{t}}$ is the cross section for the partonic
subprocesses as a function of the Mandelstam variables $\hat{s},~\hat{t},~\hat{u}$ \cite{Owens87}. The higher order corrections in the
partonic subprocesses are effectively included in (\ref{eq:eq21}) through the $K$-factor, appearing in the right hand side, which depends on the transverse momentum of the outcoming photon and the beam energy \cite{Barnafoldi01}. 

At this point we should mention that although part of the $k_T$-effects is unavoidably included in the NLO calculations, here we are studying the non-perturbative origin of such effects. To this purpose we are using a minimal modification to the standard approach in order to obtain an upper bound for such non-perturbative effects. Following this reasoning we attempt to describe experimental data introducing partonic transverse degrees of freedom through the replacement \cite{Owens87,Aurenche06}:
\begin{equation}
dx_i~f_{i/p}(x_i,Q^2) \longrightarrow dx_i d^2 k_{T,i} g(\bs{k}_{T,i}) f_{i/p}(x_i,Q^2)
\label{eq:eq22}
\end{equation}
in the PDF of the colliding partons ($i=a,b$). To avoid singularities in the partonic subprocesses we introduce a regularizing parton mass
\cite{Feynman78,Wang97} with value close to the constituent quark mass $m_q=0.3~GeV$ in the Mandelstam variables appearing in the denominator of the corresponding matrix elements. In fact $m$ can be chosen in the range $[0.1,1.0]~GeV$ without affecting the following analysis. Using the
distribution $g(k_T)$ obtained in the last section it is straightforward to calculate the cross section (\ref{eq:eq21}).

\begin{figure}
\includegraphics[width=9. cm]{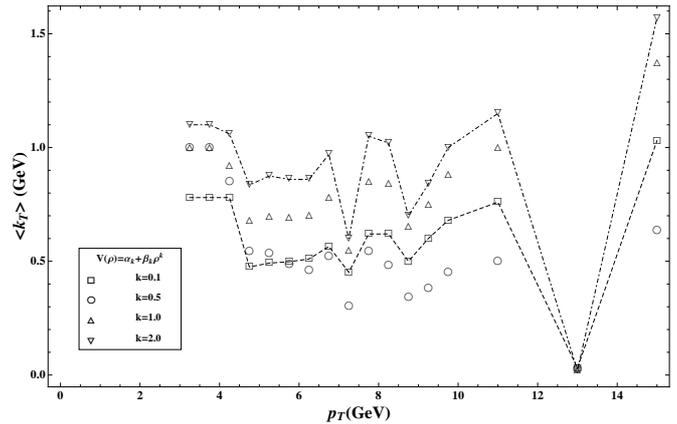}
\caption{The mean transverse momentum $\bra k_T \ket$ needed for the description of the PHENIX $pp$ differential cross section as a function of $p_T$  of the outgoing photon, using the four different partonic transverse momentum distributions shown in Fig.~\ref{fig:fig3}.}
\label{fig:fig4}
\end{figure}

We start our numerical investigations calculating the differential cross section for PHENIX data \cite{PHENIX} on prompt photon production with transverse momentum $p_T$ at RHIC ($\sqrt{s}=200~GeV$). At this step of the analysis we use the PHENIX data since, due to the very high beam energy, the $\frac{p_T}{\sqrt{s}}$ ratio is expected to become very small indicating the presence of non-perturbative QCD processes where $k_T$-effects are expected to be relevant. We perform four sets of runs, each one using a different distribution $g(k_T)$ (see Fig.~\ref{fig:fig3}) associated with the different values of $k$ ($k=0.1,0.5,1,2$) in the potential (\ref{eq:eq2}). Varying the mean transverse momentum $\bra k_T \ket$ we fit in each case all the available PHENIX data for different $p_T$ of the produced photon. As a result an one-to-one relation of $\bra k_T \ket$ with $p_T$, for each $g(k_T)$ used, is established. In Fig.~\ref{fig:fig4} we display graphically this relation for the four considered cases. In general the variations of the $\bra k_T \ket$ dependence on $p_T$ are not as large as when an ad-hoc Gaussian $g(k_T)$ is used (\cite{PRD2}). However, in order to achieve a comparison between the different models we impose the following two criteria:
\begin{itemize} 
\item compatibility of the fitted $\bra k_T \ket$-values with the geometrical properties of the proton, and
\item smoothness of the relation between $\bra k_T \ket$ and $p_T$
\end{itemize}
To make the first requirement more quantitative we calculate the $p_T$ averaged uncertainty $\sigma_{\bra k_T \ket}=\Delta \bra k_T \ket$ of 
$\bra k_T \ket$ for each of the considered models and we compare it with $\frac{\hbar}{2 r_{ch}}$ obtained using the proton charge radius shown in table I. Assuming that $\sigma_{\bra k_T \ket}$ describes successfully the $k_T$-fluctuations then the ratio:
\begin{equation} 
R_G=\frac{\sigma_{\bra k_T \ket}}{\displaystyle{\frac{\hbar}{2 r_{ch}}}}
\label{eq:eq23}
\end{equation}
can be used as a measure of the consistent description of the proton's geometry (size) within the considered model. Optimally we expect $R_G \approx 1$ while deviations may originate from the type of the inter-quark potential, the presence or not of three-body forces and the relevance or not of relativistic effects. The second criterion is quantified introducing a non-smoothness parameter $R_{NS}$ defined as the average slope variation squared in adjacent $p_T$-intervals. To be more precise one uses a linear approximation for the function $\bra k_T \ket(p_T)$, as determined by the pairs $(p_T,\bra k_T \ket)$, found through the fitting of the experimentally observed cross section for each considered model, to estimate the slope $s_i$ in the $i$-th $p_T$-interval. Then $R_{NS}$ is given by:
\begin{equation}
R_{NS}=\frac{1}{N-2} \sum_{i=2}^{N-1} (s_i - s_{i-1})^2
\label{eq:eq24}
\end{equation}
From this definition it is clear that with increasing $R_{NS}$ the associated function $\bra k_T \ket(p_T)$ becomes less and less smooth. 
The results for the quantities $R_G$ and $R_{NS}$, calculated using $g(k_T)$ obtained from the four different potential models discussed above, are summarized in Table II. For comparison we also include in this table the values of $R_G$ and $R_{NS}$ found using $g(k_T)$ determined by the MIT bag model \cite{PRD2} as well as by solving the three body problem with two-body interactions of the form (\ref{eq:eq1}). In order to be complete we give the values of these quantities found in the case of using the usual Gaussian $g(k_T)$ in the cross section calculations.\\

\begin{table}[h]
\caption{\label{tab:table2} The quantities $R_G$ and $R_{NS}$ for the considered models.}
\begin{center}
\begin{tabular}{l|cc}
Model    & $R_G$ & $R_{NS}$ \\ \hline
$k=0.1$ (3-body) & 1.27 & 0.17 \\ 
$k=0.5$ (3-body) & 1.45 & 0.18 \\
$k=1$ (3-body)   & 1.67 & 0.30 \\
$k=2$ (3-body)   & 1.88 & 0.57 \\
$k=0.1$ (2-body) & 2.97 & 1.18 \\
MIT bag & 0.85 & 0.04 \\
Gaussian & 4.06 & 3.65 \\
\end{tabular}
\end{center}
\end{table}

\begin{figure}
\includegraphics[width=9. cm]{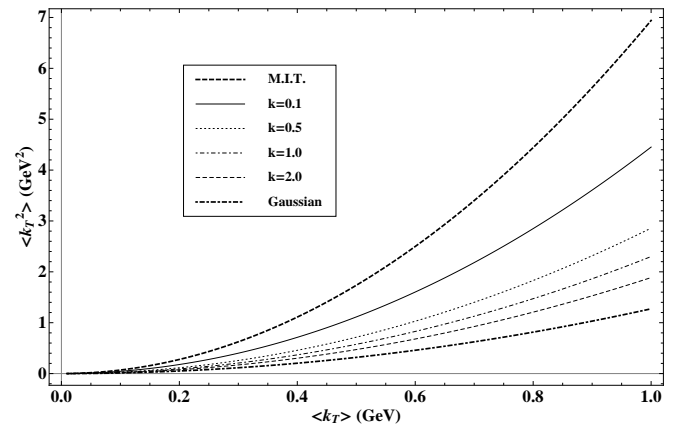}
\caption{The variance $\langle k_T^2 \rangle$ as a function of $\langle k_T \rangle$ for the considered distributions $g(k_T)$.}
\label{fig:fig7}
\end{figure}

According to Table II it is evident that the hyper-radial potential with $k=0.1$ leads to more consistent values for $R_G$ and $R_{NS}$ than the other three choices ($k=0.5$, $1.0$ and $2.0$). Clearly for $g(k_T)$ obtained from (\ref{eq:eq2}) with $k=0.1$ the fluctuations of $\bra k_T \ket$ are smaller and closer to the expectations for the proton size based on Heisenberg's uncertainty relation ($\Delta k_T \approx 0.33 GeV$). Comparing the results for the $k=0.1$ hyper-radial potential with those for the similar 2-body potential we conclude that three body forces are important for a consistent description of the partonic transverse momentum effects in the proton. In addition, the partonic transverse momentum distribution determined using the MIT bag model leads to the best values for $R_G$ and $R_{NS}$ suggesting that relativistic effects also influence significantly the transverse momentum structure of the proton. Finally, it is important to trace the behavior of each model to the characteristics of the corresponding distribution $g(k_T)$. As it is clearly seen from Fig.~\ref{fig:fig7} the best description, of the experimental data, is achieved using the model resulting in the greatest variance of $k_T$ for given $\langle k_T \rangle$ (MIT bag \cite{PRD2}). It is also interesting to notice that this characteristic depends smoothly on the exponent $k$ of the potential, becoming more pronounced for lower $k$, approaching the MIT description.

\begin{figure}
\includegraphics[width=9. cm]{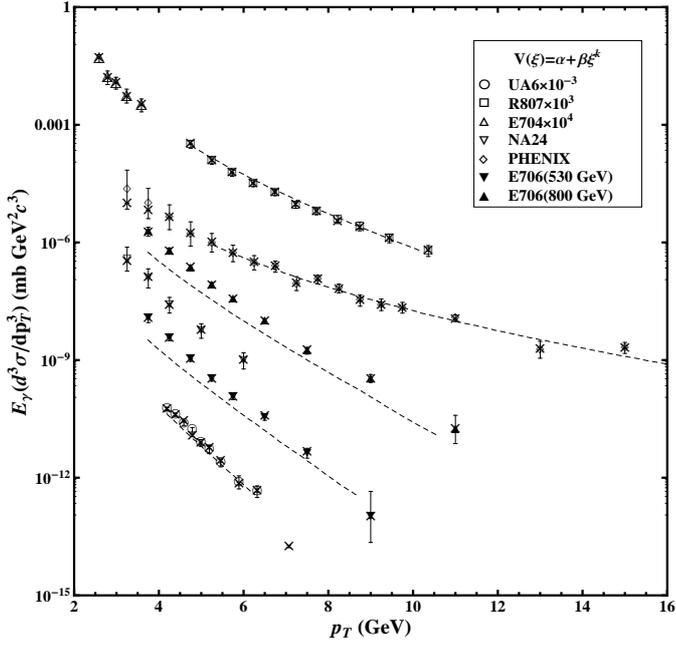}
\caption{The differential cross section for the production of a single photon with transverse momentum $p_T$ in $pp$ collisions as measured in 
various experiments. The dashed lines display results obtained using the resummation technique \cite{Sterman87,Laenen98,Catani99}. The results obtained through fitting of the data using the partonic transverse momentum distribution of Fig.~\ref{fig:fig3} with $k=0.1$ and variable $\bra k_T \ket$ practically coincide with the experimental data.}
\label{fig:fig5}
\end{figure}

\begin{figure}
\includegraphics[width=9. cm]{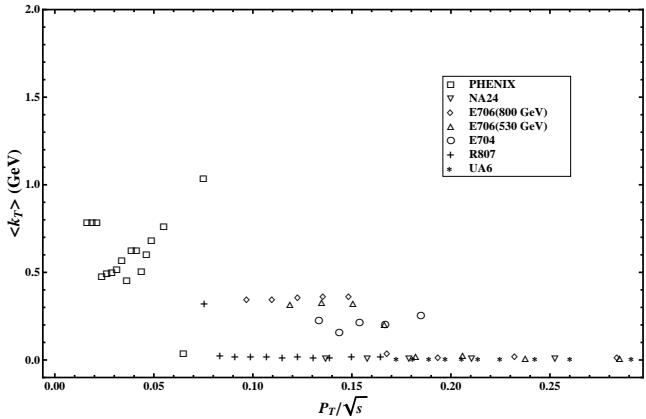}
\caption{The mean transverse momentum $\bra k_T \ket$ needed for the description of all available differential cross section data for prompt photon production in $pp$ collisions as a function of the $p_T/\sqrt{s}$ of the outgoing photon, using the $k=0.1$ partonic transverse momentum distribution shown in Fig.~\ref{fig:fig3}.}
\label{fig:fig6}
\end{figure}

Remaining in the framework of non-relativistic hyper-radial potentials it seems reasonable to restrict the detailed analysis of all existing experimental data on single $\gamma$-production in $pp$ collisions to the case of $g(k_T)$ originating from (\ref{eq:eq2}) with $k=0.1$. In Fig.~\ref{fig:fig5} we show the cross section data from 8 experiments \cite{Otherexperiments} varying both in $\sqrt{s}$ and in the observed $p_T$ region. The mean transverse momenta of the partons, necessary for a perfect description of these data is plotted in Fig.~\ref{fig:fig6}. We clearly see that only the region of small values of the ratio $\frac{p_T}{\sqrt{s}}$ requires relatively large $\bra k_T \ket$ for the data description. For $\frac{p_T}{\sqrt{s}}~>~0.05$ the necessary mean transverse momenta lie in the interval $[0,0.3]~GeV$ which is in accordance with proton's structure.

\section{Concluding remarks}

In this work we have investigated the influence of three-body forces in the transverse momentum distribution of partons inside the proton. Using a class of hyper-radial potentials (\ref{eq:eq2}) we have determined the corresponding single parton transverse momentum distribution within a non-relativistic treatment having as constraints the accurate description of the proton's ground and first excited state energy. The charge radius of the proton turns out to be almost the same ($\sim 0.6~fm$) for all potentials in the considered class. The obtained transverse momentum distributions have been incorporated in a phenomenological scheme, based on perturbative QCD, for the cross section calculation of prompt photon production in $pp$-collisions. In particular, using the associated mean transverse momentum $\bra k_T \ket$ as a free parameter, we have fitted the PHENIX cross section in a wide region of the transverse momentum of the produced photon. Within this treatment the smoothest distribution of the $\bra k_T \ket$-values, necessary for a successful description of the data, is found using the transverse momentum distribution corresponding to the potential (\ref{eq:eq2}) with $k=0.1$. This distribution has been also used for the description of all available experimental data for prompt photon production in $pp$ collisions. The $\bra k_T \ket$-spectrum leading to a perfect description of all available experimental data is found to be restricted in the range $[0.0,1.0]~GeV$. When compared with the analysis found in the literature concerning the description of the same data using a Gaussian transverse momentum distribution the results found here possess two advantages: (i) the interval of the necessary $\bra k_T \ket$-values is clearly narrower and (ii) it is displaced to smaller values which are closer to the geometrical characteristics of the proton according to Heisenberg uncertainty relation. In a similar treatment in \cite{PRD2} using the relativistic MIT bag model we have obtained an even shorter interval of $\bra k_T \ket$-values approaching the $\bra k_T \ket \approx 0$ region. Suitably defined measures for the quality of the behavior of the function 
$\bra k_T \ket(p_T)$ in the different experiments allow for a comparison between the various models and lead to the following conclusions: 
\begin{itemize}
\item In general quark confinement leaves imprint in the cross-sections for prompt photon production through the partonic transverse momentum distribution.
\item Relativistic effects are important as dictated by the results found in \cite{PRD2} using for the confinement description the MIT bag model.
\item Further study is needed in order to clarify to what extent the exact form of asymptotic freedom (the shape of the inter-quark potential for small distances) is also influencing the quality of the description of experimental data within our approach.
\item Finally it turns out that three-body forces, included in the present approach but not in the MIT bag model, are also important for an efficient
description of partonic transverse momentum effects inside the proton.
\end{itemize}  
Thus it is interesting to extend the present work by investigating the partonic transverse momentum distribution in a MIT bag model with interacting partons where three-body forces are also included. According to the findings of the present work such a model should lead to a further improvement of the description of the proton transverse momentum structure.

\begin{acknowledgement}
This work was financially supported by the Research Committee of the University of Athens (research funding program KAPODISTRIAS).
\end{acknowledgement}
{}

\end{document}